\documentclass[prl,aps,twocolumn,epsfig,superscriptaddress,showpacs]{revtex4}
\usepackage{bm}
\usepackage{amsfonts}
\usepackage[dvips]{graphicx}
\usepackage{mathrsfs}
\usepackage[intlimits]{amsmath}
\usepackage[colorlinks, citecolor=red]{hyperref}

\begin{document}

\title{Experimental Distillation of Quantum Nonlocality }
\author{C. Zu$^{1}$, D.-L. Deng $^{1,2}$, P.-Y. Hou$^{1}$, X.-Y. Chang$^{1}$%
, F. Wang$^{1}$, L.-M. Duan}
\affiliation{Center for Quantum Information, IIIS, Tsinghua University, Beijing 100084,
PR China}
\affiliation{Department of Physics, University of Michigan, Ann Arbor, Michigan 48109, USA}
\date{\today}

\begin{abstract}
We report the first experimental demonstration of distillation of quantum
nonlocality, confirming the recent theoretical protocol [\textit{Phys. Rev.
Lett. 102, 120401 (2009)}]. Quantum nonlocality is described by a
correlation box with binary inputs and outputs, and the nonlocal boxes are
realized through appropriate measurements on polarization entangled photon
pairs. We demonstrate that nonlocality is amplified by connecting two
nonlocal boxes into a composite one through local operations and four-photon
coincidence measurements.
\end{abstract}

\pacs{03.65.Ta, 42.50.Xa, 03.65.Ud,03.65.Ca}
\maketitle

The seminal paper by Bell in $1964$ reveals that spatially separated quantum
systems could have wield correlation, impossible to be explained by any
local hidden variable (LHV) theory based on shared randomness \cite%
{1964Bell,1969Clauser}. This strong correlation is known thereafter as
quantum nonlocality, which has been tested by a number of remarkable
experiments \cite{1982Aspect,1998Weihs,2001Rowe}. Quantum nonlocality is not
only the critical concept for foundational research of quantum mechanics,
but also finds important applications in recent development of quantum
information theory. Nonlocality emerges as a key resource, different from
entanglement, for realization of various quantum information protocols, such
as device-independent quantum key distribution~\cite%
{1991Ekert,2005Barrett,2007Acin}, nonlocal computation~\cite{2007Linden},
and self-certified random number generators \cite{2007Colbeck,2010Pironio}.

Similar to entanglement, nonlocal correlation is more useful if it gets
stronger. Entanglement purification protocols have been proposed \cite%
{1996Bennett} and demonstrated by several experimental groups \cite%
{2001Kwiat,2003Pan,2006Reichle}. An interesting question is whether
nonlocality can be distilled. Can we get stronger nonlocality from local
operations on multiple weakly nonlocal systems? The answer is far from being
obvious as entanglement purification does not automatically yield
nonlocality distillation. Entanglement purification protocols in general use
both local operation and classical communication \cite{1996Bennett}, while
classical communication is not allowed for distillation of nonlocality as it
violates the locality requirement. Note that entanglement and nonlocality
characterize essentially different resources \cite{2005Brunner}. It is well
known that there are entangled states where the quantum correlation can be
described by the LHV\ theory with no nonlocality \cite{1989Werner}. Because
of this difference, it has been proven that for a large class of
nonlocality, distillation is actually impossible \cite{2009Forster, ND}.
Only until recently theoretical advance finds examples to show that certain
nonlocality described by correlation boxes can be distilled through only
local operations \cite{2009Forster,2009Brunner}.

In this paper, motivated by this intriguing theoretical advance \cite%
{2009Forster,2009Brunner}, we report the first experimental demonstration of
quantum nonlocality distillation using the photonic system. By controlling
the bases of binary measurements on entangled photon pairs, we realize the
nonlocal boxes proposed by Forster et al. \cite{2009Forster} that allow
distillation of nonlocality. The nonlocality is quantified by its violation
of the Clauser-Horne-Shimony-Holt (CHSH) inequality \cite{1969Clauser}. We
optimize the experimental parameters to maximize the nonlocality difference
between the distilled box and the original box. The difference is typically
small (about a few percents) even for the optimized box, so the experiment
needs to be precisely controlled. To measure correlation and nonlocality of
the distilled box formed with local operations on two identical nonlocal
boxes, we use linear optics elements and four-photon coincidence detection,
and the experimental data unambiguously demonstrate nonlocality increase of
the distilled box compared with the original individual ones.

The nonlocality revealed by violation of Bell's inequality can be described
by a correlation box shared between two parties, typically called Alice and
Bob \cite{1994Popescu}. We consider a Bell scenario with binary inputs and
outputs. Each party has an input bit, denoted by $x,y\in \{0,1\}$, that
determines his/her measurement basis, and an output bit, denoted by $a,b\in
\{0,1\}$, that corresponds to the measurement outcome. The conditional
probability $\mathtt{P}(ab|xy)$ completely determines the correlation of the
box. Under a given basis $\left( x,y\right) $, the correlation of the
measurement outcomes $\left( a,b\right) $ is described by
\begin{equation}
\mathtt{\mathbf{C}}_{xy}(\mathtt{P})=\mathtt{P}(00|xy)+\mathtt{P}(11|xy)-%
\mathtt{P}(01|xy)-\mathtt{P}(10|xy).
\end{equation}%
From this correlation function, one can define the CHSH nonlocality of the
box \cite{1969Clauser,2009Forster,2009Brunner}, which is characterized by
\begin{equation}
\mathcal{N}(\mathtt{P})=\max_{xy}|\mathtt{\mathbf{C}}_{xy}(\mathtt{P})+%
\mathtt{\mathbf{C}}_{x\bar{y}}(\mathtt{P})+\mathtt{\mathbf{C}}_{\bar{x}y}(%
\mathtt{P})-\mathtt{\mathbf{C}}_{\bar{x}\bar{y}}(\mathtt{P})|.
\end{equation}%
The algebraic maximum of $\mathcal{N}(\mathtt{P})$ is $4$, however, $%
\mathcal{N}(\mathtt{P})$ needs to be bounded by tighter values for different
physical theories, and the value of $\mathcal{N}(\mathtt{P})$ characterizes
the maximum nonlocality achievable in such a theory. The LHV\ theory based
on the assumption of local realism requires that the correlation in $\mathtt{%
P}(ab|xy)$ is form pre-shared randomness, that is, $\mathtt{P}(ab|xy)$ can
be written in the form $\mathtt{P}(ab|xy)=\int \mathtt{P}(a|x,v)\mathtt{P}%
(b|y,v)p\left( v\right) dv$, where $v$ is the hidden random variable with
the probability distribution $p\left( v\right) $. With this restriction on
the correlation in $\mathtt{P}(ab|xy)$, the well known CHSH\ inequality
shows that the nonlocality measured by $\mathcal{N}(\mathtt{P})$ is bounded
from above by $\mathcal{N}(\mathtt{P})\leq 2$ for any models based on local
realism \cite{1969Clauser}. Quantum mechanics allows stronger correlation,
and any violation of the CHSH inequality with $\mathcal{N}(\mathtt{P})>2$ is
a signature of nonlocality. However, nonlocality in quantum mechanics is
still bounded by the Tsirelson bound with $\mathcal{N}(\mathtt{P})\leq 2%
\sqrt{2}$ \cite{1980Tsirelson-bound}. It is interesting to note that the
non-signalling condition alone from the relativity theory in principle could
allow even stronger nonlocality. In term of the correlation matrix $\mathtt{P%
}(ab|xy)$, the non-signalling condition requires the marginal distribution $%
\mathtt{P}(a|xy)\equiv \sum_{b}\mathtt{P}(ab|xy)=\mathtt{P}(a|x)$,
independent of $y$, and $\mathtt{P}(b|xy)\equiv \sum_{a}\mathtt{P}(ab|xy)=%
\mathtt{P}(b|y)$, independent of $x$. This condition guarantees that two
remote parties can not signal (change the marginal distribution of the other
side) by choosing different measurement bases. With the non-signaling
condition alone, Popescu and Rohrlich (PR) have constructed a nonlocal box,
the so-called PR box, that achieves the maximum algebraic violation of the
CHSH inequality with $\mathcal{N}(\mathtt{P})=4$ \cite{1994Popescu}. The
nonlocality in the range of $2\sqrt{2}<\mathcal{N}(\mathtt{P})\leq 4$,
although not attainable by quantum mechanics, can be discussed in the
general framework of non-signalling theory \cite%
{1994Popescu,2006Masanes,2009Forster,2009Brunner}.

To distill nonlocality shared between two parties using only local
operations, we consider a particular type of nonlocal correlation boxes
proposed in Ref. \cite{2009Forster}, for which the conditional probability
matrix $\mathtt{P}(ab|xy)$ is parametrized in the following way

\begin{eqnarray}
\mathtt{P} &\equiv &\left(
\begin{matrix}
\mathtt{P}(00|00) & \mathtt{P}(01|00) & \mathtt{P}(10|00) & \mathtt{P}(11|00)
\\
\mathtt{P}(00|01) & \mathtt{P}(01|01) & \mathtt{P}(10|01) & \mathtt{P}(11|01)
\\
\mathtt{P}(00|10) & \mathtt{P}(01|10) & \mathtt{P}(10|10) & \mathtt{P}(11|10)
\\
\mathtt{P}(00|11) & \mathtt{P}(01|11) & \mathtt{P}(10|11) & \mathtt{P}(11|11)%
\end{matrix}%
\right)  \notag \\
&=&\frac{1}{2}\left(
\begin{matrix}
1-\eta & \eta & \eta & 1-\eta \\
1-\eta & \eta & \eta & 1-\eta \\
1-\eta & \eta & \eta & 1-\eta \\
1-\gamma & \gamma & \gamma & 1-\gamma%
\end{matrix}%
\right) ,
\end{eqnarray}%
with $0<\eta ,\gamma <1$. It is easy to check that the CHSH nonlocality for
this matrix is given by $\mathcal{N}(\mathtt{P})=2+2\gamma -6\eta $. This
correlation box is nonlocal when $\gamma >3\eta $. However, not every box is
realizable with a physical system, even when $\mathcal{N}(\mathtt{P})\leq 2%
\sqrt{2}$ which satisfies the Tsirelson bound. A necessary and sufficient
condition for a set of correlation functions $\mathtt{\mathbf{C}}_{xy}(%
\mathtt{P})$ to be attainable by quantum mechanics has been derived in Refs.~%
\cite{1988Landau,2003Masanes}, which implicitly determine the physical
region of $\eta ,\gamma $ \cite{2009Forster}. For two nonlocal boxes
characterized by the same conditional probability matrices $\mathtt{P}%
(a_{1}b_{1}|xy)$ and $\mathtt{P}(a_{2}b_{2}|xy)$ in the form of Eq. (3) with
the same input $x,y$ for the bases of detection but different measurement
outcomes $a_{1},b_{1}$\ and $a_{2},b_{2}$, the local distillation operation
is done through a mod $2$ addition of the measurement outcomes on each side
as illustrated in Fig. 1, that is, the distilled box is characterized by the
condition probability $\mathtt{P}_{d}(ab|xy)$, with $a=a_{1}\oplus a_{2}$
and $b=b_{1}\oplus b_{2}$ \cite{2009Forster}. It is easy to check that the
matrix for $\mathtt{P}_{d}(ab|xy)$ still has the form of Eq. (3), but with $%
\eta ,\gamma $ replaced by $\eta ^{\prime },\gamma ^{\prime }$, where%
\begin{equation}
\eta ^{\prime }=2\left( \eta -\eta ^{2}\right) ,\text{ \ }\gamma ^{\prime
}=2\left( \gamma -\gamma ^{2}\right) .
\end{equation}%
The distilled box has stronger nonlocality compared with the original box if
$\mathcal{N}(\mathtt{P}_{d})>\mathcal{N}(\mathtt{P})$. A necessary condition
for this is that the parameters $\eta ,\gamma $ are in the region $0<\eta
<\gamma /3<1/6$. For experimental implementation of the nonlocality
distillation, it is better to have $\mathcal{N}(\mathtt{P}_{d})-\mathcal{N}(%
\mathtt{P})$ as large as possible. Under the constraint that the box
characterized by the conditional probability in the form of Eq. (3) is
physically attainable, we numerically maximize the nonlocality increase $%
\mathcal{N}(\mathtt{P}_{d})-\mathcal{N}(\mathtt{P})$ under different
parameters $\eta ,\gamma $ and find that the optimal values are $\eta
_{o}\approx 0.019$ and $\gamma _{o}\approx 0.164$. Under this optimal choice
of $\eta ,\gamma $, the nonlocality increase $\mathcal{N}(\mathtt{P}_{d})-%
\mathcal{N}(\mathtt{P})\approx 2.324-2.214=0.110$, representing about a $5\%$
improvement. As the relative increase in nonlocality is small, the
experiment needs to be done with a good precision for an unambiguous
demonstration of nonlocality distillation.

To experimentally realize distillation of two nonlocal boxes, we first need
to implement a correlation box where the conditional probability $\mathtt{P}%
(ab|xy)$ has the form of Eq. (3) with tunable $\eta ,\gamma $. We assume
Alice and Bob share singlet entangled states given by $|\psi ^{-}\rangle
=(|01\rangle -|10\rangle )/\sqrt{2}$, and Alice (Bob) measures a Paul spin $%
\mathbf{\sigma }_{A}$ ($\mathbf{\sigma }_{B}$) along the $\mathbf{n}_{0},%
\mathbf{n}_{1}$ ($\mathbf{m}_{0},\mathbf{m}_{1}$) direction when the input
bit $x=0,1$ ($y=0,1$). The output bit is taken as $a=0,1$ ($b=0,1$) if the
measurement outcome of the Paul spin $A_{x}\equiv \mathbf{n}_{x}\cdot
\mathbf{\sigma }_{A}$ ($B_{y}\equiv \mathbf{m}_{y}\cdot \mathbf{\sigma }_{B}$%
, $x,y=0,1$) is $+1,-1$, respectively. In this implementation, one can check
that the conditional probability $\mathtt{P}(ab|xy)$ has the form of Eq. (3)
if $\mathbf{n}_{0}\cdot \mathbf{m}_{0}=\mathbf{n}_{0}\cdot \mathbf{m}_{1}=%
\mathbf{n}_{1}\cdot \mathbf{m}_{0}$, with $\eta =(1+\mathbf{n}_{0}\cdot
\mathbf{m}_{0})/2$ and $\gamma =(1+\mathbf{n}_{1}\cdot \mathbf{m}_{1})/2$.
To satisfy this constraint, we take the directions of $\mathbf{n}_{0},%
\mathbf{n}_{1},\mathbf{m}_{0},\mathbf{m}_{1}$ as specified by the angle $%
\varphi $ in Fig. 2. In this case, $\mathbf{n}_{0}\cdot \mathbf{m}_{0}=%
\mathbf{n}_{0}\cdot \mathbf{m}_{1}=\mathbf{n}_{1}\cdot \mathbf{m}_{0}=-\cos
\left( \varphi \right) $ and $\mathbf{n}_{1}\cdot \mathbf{m}_{1}=-\cos
\left( 3\varphi \right) $. We find that with $\varphi =15.95^{o}$, this
implementation realizes the optimal choice of $\eta ,\gamma $ with $\eta
=(1-\cos \left( \varphi \right) )/2\approx 0.019$ and $\gamma =(1-\cos
\left( 3\varphi \right) )/2\approx 0.164$ that maximize the nonlocality
increase for distillation of two nonlocal boxes.

To realize the two nonlocal correlation boxes specified with the above
conditions, we experimentally generate two pairs of entangled photons
through the spontaneous parametric down-conversion (SPDC) setup shown in
Fig. 3. Two pieces of the type-II BBO crystals are pumped by femtosecond
laser pulses, generating two pairs of entangled photons in the singlet state
$|\psi ^{-}\rangle =(|HV\rangle -|VH\rangle )/\sqrt{2}$ \cite{21}, where $%
|H\rangle $ and $|V\rangle $ denote horizontal and vertical polarization of
a single photon. After the entangled photons are generated, Alice (Bob)
applies the measurement $A_{x}$ ($B_{y}$) by using half-wave plates (HWP) to
rotate the polarization of her (his) photons. The angles for the wave plates
HWP5, HWP6, HWP7, HWP8 are specified in the supplementary information
corresponding to the four different inputs $\left( x,y\right) =\left(
0,0\right) ,\left( 0,1\right) \left( 1,0\right) \left( 1,1\right) $ of the
correlation box. For the nonlocal boxes 1 and 2, the measurement outcomes
for the conditional probabilities $\mathtt{P}_{1}(a_{1}b_{1}|xy)$ and $%
\mathtt{P}_{2}(a_{2}b_{2}|xy)$ are recorded in the table of Fig. 4. From the
measurements, we find the CHSH nonlocality $\mathcal{N}\left( \mathtt{P}%
_{1}\right) =2.1440\pm 0.0001$ and $\mathcal{N}\left( \mathtt{P}_{2}\right)
=2.1356\pm 0.0001$, where the error bar accounts for the statistical error
associated with the photon counts under the assumption of a Poissonian
distribution. The realized two nonlocal boxes are close to the optimal box
specified above for distillation. The small difference is due to the
infidelity of the entangled singlet states as well as the imprecision in
controlling the angles of the wave plates.

To measure the conditional probabilities $\mathtt{P}_{d}(ab|xy)$ for the
distilled box realized in Fig. 3, we note that the mod $2$ addition $%
a=a_{1}\oplus a_{2}$ and $b=b_{1}\oplus b_{2}$ required in the distillation
protocol can be simply implemented with a polarization beam splitter (PBS).
A PBS transmits (reflects) the photon when it is in $H$\ ($V$) polarization.
The outputs of the PBS are coupled into single mode fibers and detected by
single photon detectors for measurement of coincidence. When we register a
four-photon coincidence between the two output modes of the PBS\ on both
Alice's and Bob's sides, the photons at the two input modes of each PBS must
have identical polarization, which means $a=a_{1}\oplus a_{2}=0$ and $%
b=b_{1}\oplus b_{2}=0$. The count rate of this coincidence is therefore
proportional to the conditional probability $\mathtt{P}_{d}(a=0,b=0|xy)$. To
measure other components of $\mathtt{P}_{d}(a,b|xy)$, we rotate the HWP9
(HWP10) at Alice's (Bob's) side by $45^{o}$, which exchanges $H$\ and $V$
and thus flips $a_{1}$ ($b_{1}$). With a bit flip on $a_{1}$, $b_{1}$, or
both, the coincidence measures the relative conditional probabilities $%
\mathtt{P}_{d}(a=1,b=0|xy)$, $\mathtt{P}_{d}(a=0,b=1|xy)$, and $\mathtt{P}%
_{d}(a=1,b=1|xy)$, respectively. A technical problem for this measurement is
that the four-photon coincidence could also be caused by the events with two
entangled photon pairs from the same BBO\ crystal and no photon from the
other crystal \cite{22}. To deduct the coincidence due to these unrelated
events, we measure their contribution to the four-photon coincidence rate
simply by blocking the down converted photons from one of the BBO\ crystals.
After this correction, the four-photon coincidence rate is directly
proportional to the conditional properties $\mathtt{P}_{d}(a,b|xy)$ for the
distilled box.

The measured conditional probabilities $\mathtt{P}_{d}(a,b|xy)$ for the
distilled box are shown in the table of Fig. 4. From these data, we find the
CHSH nonlocality $\mathcal{N}(\mathtt{P}_{d})=2.206\pm 0.021$. The error bar
gets larger since to measure the properties of the distilled box we need to
record four-photon coincidence, which has a significantly smaller count rate
and thus a larger statistical error. Apparently, $\mathcal{N}(\mathtt{P}%
_{d})>\mathcal{N}\left( \mathtt{P}_{1}\right) $ and $\mathcal{N}(\mathtt{P}%
_{d})>\mathcal{N}\left( \mathtt{P}_{2}\right) $, where the nonlocality
increases by more than three times the standard deviation (error bar), so
the experiment unambiguously demonstrate distillation of quantum nonlocality.

In summary, we have reported the first experimental demonstration of
distillation of quantum nonlocality through only local operations on two
correlation boxes. From a fundamental point of view, the experiment
unambiguously confirms that the wield correlation of quantum mechanics, the
nonlocality unexplainable by any local realistic theory, can be enhanced
without any communication (quantum or classical) between the remote parties.
From a practical point of view, nonlocality has emerged as an important
resource for implementation of self-certified device-independent quantum
information protocols, and an experimental demonstration of nonlocality
amplification provides a useful step for future applications along this line.

\textbf{Acknowledgement} This work was supported by the National Basic
Research Program of China (973 Pro- gram) 2011CBA00300 (2011CBA00302) and
the NSFC Grant 61033001. DLD and LMD acknowledge in addition support from
the IARPA MUSIQC program, the ARO and the AFOSR MURI program.

\begin{figure}[tbp]
\includegraphics[width=42mm]{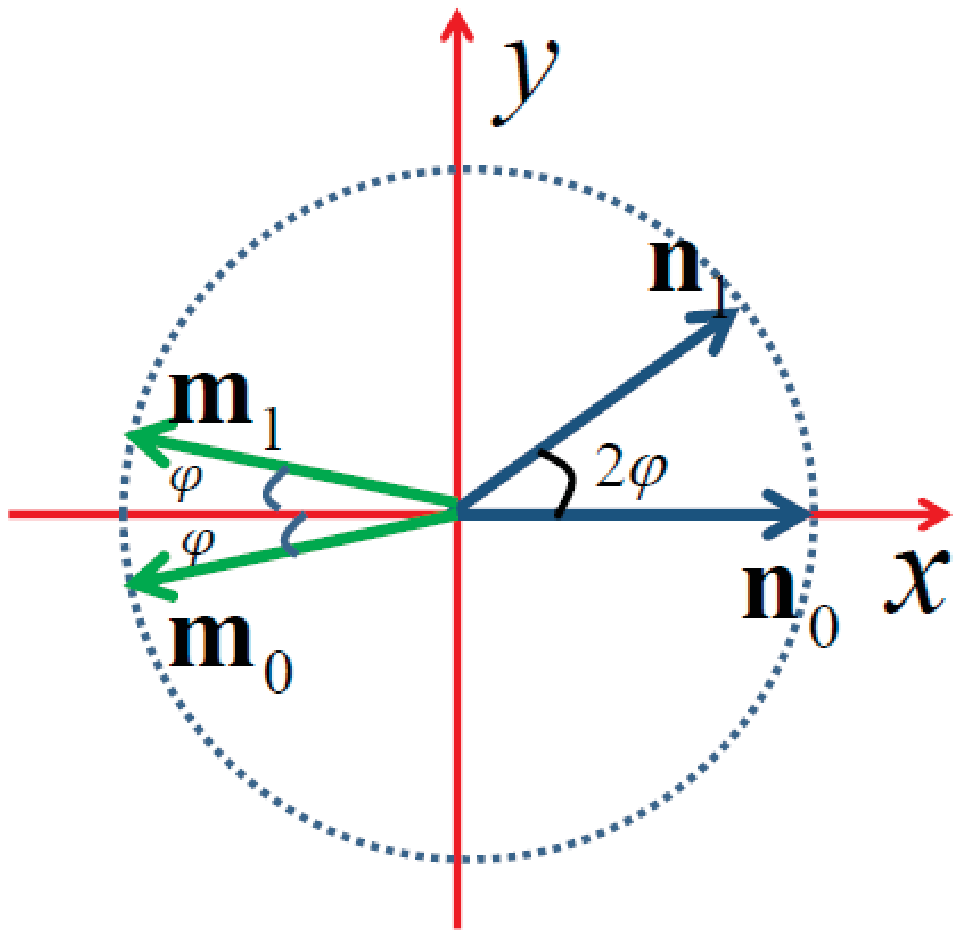}\newline
\caption{(Color online) The directions of the measurements of the Pauli
spins for Alice ($\mathbf{n}_0,\mathbf{n}_1$) and Bob ($\mathbf{m}_0,\mathbf{%
m}_1$), which realize the correlation box characterized by the conditional
probabilities in the form of Eq. (3) that is optimal for demonstration of
nonlocality distillation from two copies.}
\label{Measurement-Settings}
\end{figure}

\begin{figure}[tbp]
\centering
\includegraphics[width=8.5cm,height=6.5cm]{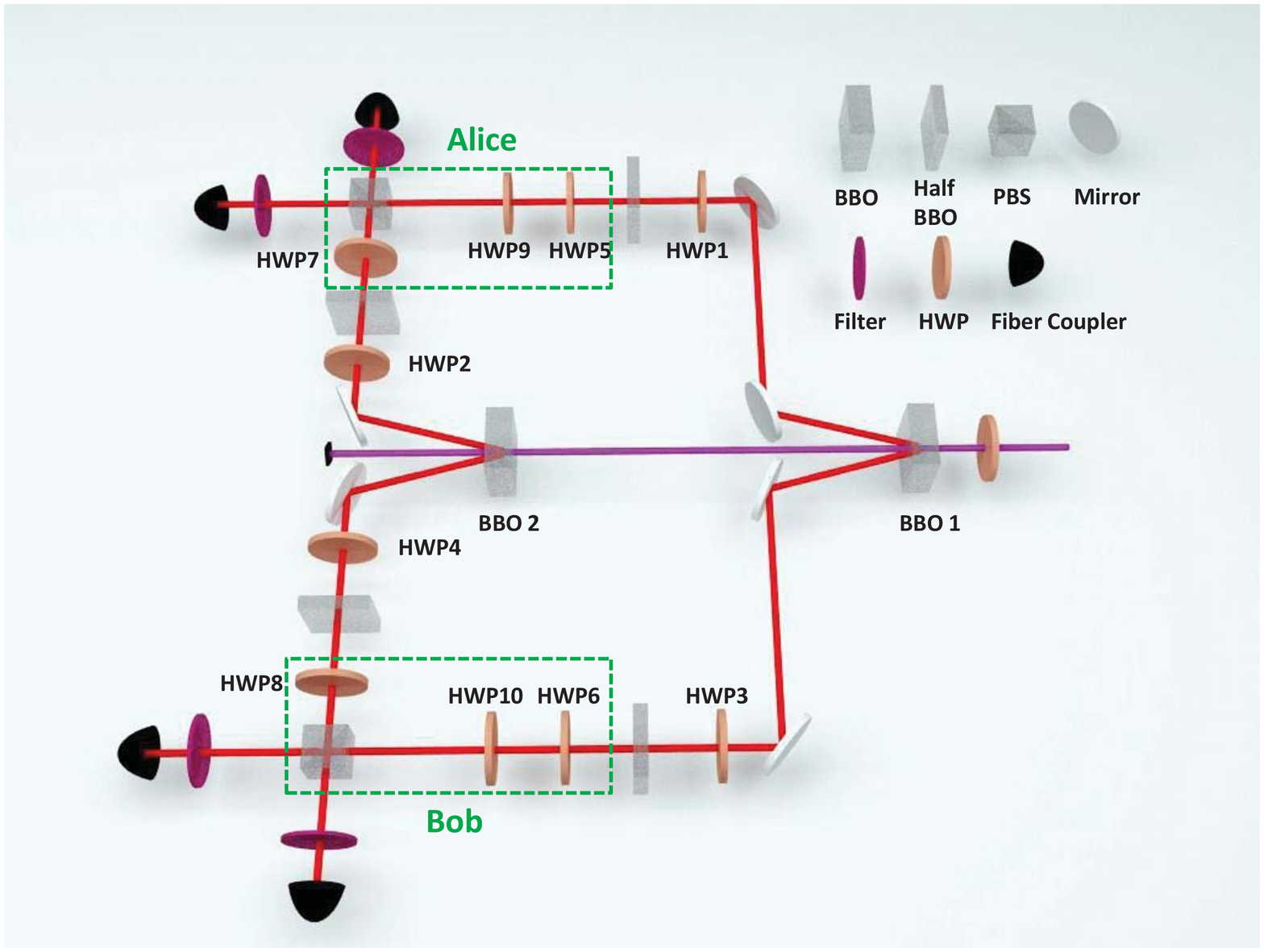}
\caption[Fig. 3 ]{The schematic experimental setup to implement the
nonlocality distillation and to measure the properties of the distilled box.
Femtosecond pulses (with the wavelength at 390nm and a repetition rate of $%
76 $ MHz) from a frequency-doubled Ti:Sa laser pump two BBO crystals (with
type-II cutting of $2$ mm depth) to generate two pairs of photons with
perpendicular polarization. Four additional BBO crystals of $1$ mm depth are
used to compensate the spatial and temporal walk-off between the photons,
which, together with the four half-wave plates (HWP1, HWP2, HWP3, HWP4) set
at $45^o$, prepare the two pairs of photons each in the maximally entangled
singlet state $|\protect\psi^-\rangle=\frac{1}{\protect\sqrt{2}}%
(|HV\rangle-|VH\rangle)$. Alice and Bob then use rotation of HWP5, HWP6,
HWP7, and HWP8 to choose the measurement bases. By measuring the spins along
the directions of $\mathbf{n}_0,\mathbf{n}_1,\mathbf{m}_0,\mathbf{m}_1$
specified in Fig. 2, Alice and Bob realize two correlation boxes which allow
maximum distillation of nonlocality using the protocol illustrated in Fig.
1. The two polarization beam splitters (PBSs) at Alice's and Bob's sides,
together with HWP9 and HWP10, realize effectively the required mod-$2$
addition. The output modes of the PBSs are coupled into single-mode fibers
and then detected by four single-photon detectors. The results are
registered through a four-port coincidence circuit with a $3$ ns coincidence
window to reduce the accidental coincidence counts. There is no need of
background subtraction of accidental coincidences for this experiment. The
typical two-photon coincidence rate from each BBO crystal is about $15kHz$
and the four-photon coincidence rate is about $2.2Hz$ for this experiment.
To reduce the statistical error, we accurate the photon counts for $9.3$
hours for each data point.}
\end{figure}

\begin{widetext}

\begin{figure}
\includegraphics[width=180mm]{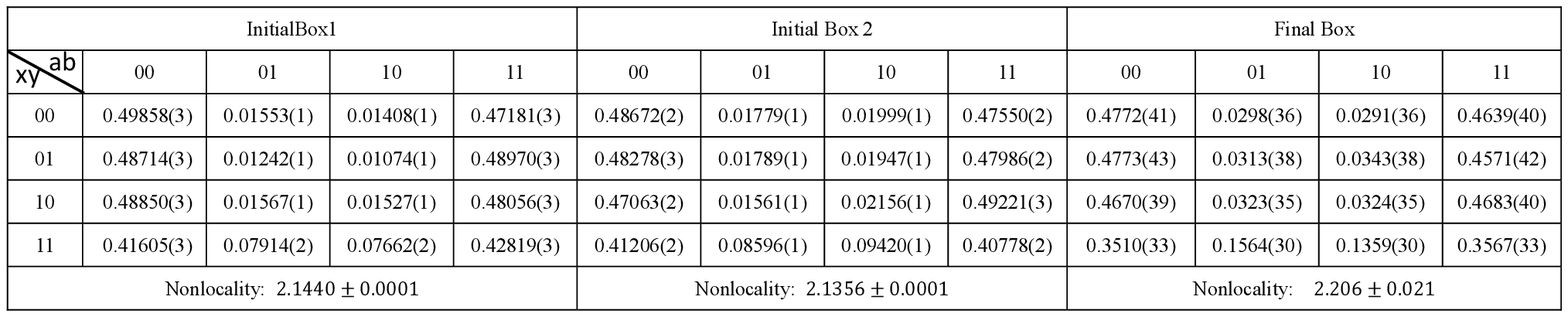}\\
 \caption{(Color online) Table I| The experimental results for the conditional probabilities
 $P(a_1b_1|xy)$ and $P(a_2b_2|xy)$ measured for each individual nonlocal box, and $P(ab|xy)$ measured
 for the distilled box. The CHSH nonlocality is calculated for each box from the measured conditional probabilities.
 The distilled box has a nonlocaity larger than that of each individual box, confirming distillation of the nonlocality.
 }\label{Exp-ResTable}
\end{figure}
\end{widetext}

\end{document}